\documentclass[12pt]{article}

\usepackage{amsthm}
\usepackage{extmath}
\usepackage{latexsym}

\newcommand{\be}{\begin{equation}}
\newcommand{\ee}{\end{equation}}
\newcommand{\ba}{\begin{eqnarray}}
\newcommand{\ea}{\end{eqnarray}}

\theoremstyle{plain}
\newtheorem{tm}{Theorem}
\newtheorem{pr}[tm]{Proposition}

\theoremstyle{remark}
\newtheorem{rk}[tm]{Remark}

\theoremstyle{definition}
\newtheorem{df}[tm]{Definition}

\begin{document}

\title{\bf Stochastic quantum process}

\author{ {\bf Jerzy Stry\l{}a \footnote{
e-mail address: jstryla@wp.pl; http://www.wnow.net/jspage } } }

\date{}

\maketitle

\begin{abstract}
I propose to treat quantum evolution as a stochastic process
consisting from a sequence of doubly stochastic matrices, which naturally
arise in the generalized quantum evolution. Then it is proved that
the law of non-decreasing entropy is fulfilled and that the law
characterizes doubly stochastic matrices. Finally, an application of the model
to support the generalized second law of black hole thermodynamics and a relation to the quantum histories formulation of quantum physics appear.

\end{abstract}

I focus in the article on two successful physical theories, quantum unitary
and statistical physics to find a relation between them, see ref.
\cite{tH99}. The motivation is that the observed law of non-decreasing
entropy in all physical processes can not be directly modeled in the
invertible treating. In classical mechanics the analogue trouble was solved
by admitting collision points of evolution in H equation.
In quantum physics irreversibility seems to be caused by measurements, a
discontinuity of unitary evolution \cite{Pa89}. The nature of measurement is
not comprehended, for a mathematical model with use of Ito calculus for
Schrodinger equation on reduced Hilbert space see \cite{BH01}, while its
functional description is core of contemporary quantum theories. I do not
solve the measurement problem, but extract all we know about it and use in
wider context. Namely, I further
assume that measurements belong to physical phenomena and they happen
independently on conscious beings and the role of measuring devices
participating in the process. For a quantum system interacting with an
environment with qualitatively higher degree of freedom I assume that
approximately point events of reductions violating unitarity take places.
Precisely, in the place of unitary evolution I put transformation of
quantum states represented by rays
via Kraus representation, so I admit a nearest quantum surrounding
\cite{LBW00,Sc00}. Whole environment influences the system in the way
that reductions to pure states appear, compare \cite{Pa89}.
The reduction itself is
treated as a Poisson process of time events.
This assumption of the model refers to
uniform conditions the system is subjected to. It will appear that it is not
essential for truthfulness of the non-decreasing entropy theorem.

Possessing the sequence of reduction moments for one entity $t_0< t_1 <
t_2< ...$ the quantum evolution describes the steps from
$t_i$ to $t_{i+1}$. In Feynman integrals approach it means that admitted
quantum trajectories need to cross only pure states
in the pointed time moments $t_i$.
For a finite dimensional Hilbert space ${\cal H}$ the skips are
realized by stochastic matrices $M\in M(n\times n, {\bbold R})$ acting on
${\bbold R}^n$, where a complete set of observables (an orthonormal basis
in the Hilbert space) was chosen. The normalized states are represented by
points of $\triangle _{n-1}\subset {\bbold R}^n$, $\triangle
_{n-1}:=\{ \{p_i\} | p_i\ge 0 \wedge \sum p_i=1 \} $. Really, the $\{p_i \}$
corresponds to a
ray in ${\cal H}$ which is sufficient subject of quantum evolution, for
example see ref. \cite{AS97}.


It will appear that only special stochastic matrices play a role in the
quantum context, see Land\'e's conjencture \cite{Lo97}.
Let us remind their definition.
\begin{df}
Let $M$ be a stochastic matrix i.e. $M\in M(n\times n,{\bbold R})$,
$M_{ij}\ge 0$, $\sum_{i=1}^{n}M_{ij}=1$. Then $M$ is doubly stochastic (DS)
iff $\sum_{j=1}^n M_{ij}=1$.
\end{df}

Stochastic matrices appear as descending objects from unitary
evolution in presence of a quantum surrounding
between two different time moments. They
are constructed via a Kraus representation. Then in the
finite dimensional case, I study in the article, a finite set of operators
$A_{\alpha}\in L({\bbold C}^n, {\bbold C}^n), \alpha =1,...,N $ exists,
which are built by contraction of an unitary higher dimensional dynamics,
such that
\be       \label{Kraus}
M_{ij}=\sum_{\alpha=1}^{N} Tr(P_iA_{\alpha}P_jA_{\alpha}^{*})
\ee
and
\be     \label{1}
\sum_{\alpha=1}^{N}  A_{\alpha}^*A_{\alpha}=1=\sum_{\alpha=1}^{N}
A_{\alpha}A_{\alpha}^*,
\ee
where $P_i$ are the projection operators define by the standard orthonormal
basis $\{|i>\} \in {\bbold R}^n\subset {\bbold C}^n$. Using $\sum P_i=id$
and (\ref{Kraus}),(\ref{1}) one concludes that such $M$ is a doubly
stochastic matrix. From the other side having any finite doubly stochastic
matrix $M$ acting on ${\bbold R}^n$ to find a representation it is enough
to put $A_{i,j}=|i>\sqrt{ M_{ij} }<j|$ as an element of $L({\bbold
C}^n,{\bbold C}^n)$. The doubly stochastic condition guarantees fulfilling
(\ref{1}).

For ordinary unitary evolution transformation to stochastic matrices, given
by $M_{ij}=|<i|U|j>|^2$ and $A_1=U\in U(n,{\bbold C}), N=1$, does not cover
all doubly stochastic matrices for $n>2$. Nevertheless all doubly
stochastic matrices, so admitting a Kraus representation with condition
(\ref{1}), may be build from a higher dimensional unitary evolution. I show

\begin{pr}
Let a finite set $\{ A_{\alpha } \}_{\alpha =1}^N$ be given, $A_{\alpha}\in
L({\bbold C}^n,{\bbold C}^n) $ and fulfills (\ref{1}). Then there is a
finite dimensional complex Hilbert space ${\cal H}_2$, $dim{\cal H}_2\le
2N$ and an unitary operator on ${\cal H}_1\otimes {\cal H}_2 $, ${\cal
H}_1:={\bbold C}^n$, such that the reduction to ${\cal H}_1$ leads to
$M(A_{\alpha})$ defined by (\ref{Kraus}).
\end{pr}

\begin{proof}
Let $I=\{1,...,N \}$ be a set of the indices of the operators $A_{\alpha}$. I
put ${\cal H}_2={\bbold C}^N\oplus {\bbold C}^N$. Then ${\cal H}_1\otimes
{\cal H}_2\simeq {\cal H}_1^I\oplus{\cal H}_1^I$. At the beginning I define
a seeking unitary operator firstly only on the diagonal of ${\cal H}_1^I$,
$\triangle _I$ by

\be
U(|v>\oplus \ldots \oplus |v> ):=\sqrt{N} A_1|v>\oplus \ldots \oplus
\sqrt{N}A_N|v>, |v>\in {\cal H}_1.
\ee

Let ${\cal H}_1^I=\triangle _I\oplus \triangle _I^{\bot} $ and ${\cal
H}_1^I=U(\triangle _I)\oplus U(\triangle_I)^{\bot }$ be the orthogonal
decompositions of the Hilbert space. An isometry $m: \triangle_I^{\bot
}\mapsto U(\triangle_I)^{\bot }$ may be chosen. And then the complete
definition is as follows

\ba
U|\triangle_I \oplus \triangle _I:=U|\triangle _I\oplus U|\triangle_I \\
U:\triangle^{\bot}_I \oplus \{ 0 \} \ni |v> \oplus \ 0 \mapsto 0 \oplus
m(|v>) \in
\{0\}\oplus U(\triangle_I)^{\bot}
\\ U:\{0\}\oplus \triangle_I^{\bot} \ni 0 \oplus |v>
\mapsto m(|v>)\oplus \in U(\triangle_I)^{\bot } \oplus \{0\}
\ea

Then the starting $M$ is rebuilt by
$M_{ij}=\frac{1}{2N}\sum_{\alpha,\sigma}|<i\alpha \sigma|U|j \alpha
\sigma>|^2$, where $|i,\alpha ,\sigma >=|i>\otimes (|\alpha >\otimes |\sigma >)$; $i,j=1,\ldots ,n;\alpha\in I;\sigma=+,-$, is the orthonormal basis of
${\cal H}_1\otimes {\cal H}_2$.
\end{proof}


It has just appeared that DS matrices play
additionally essential role while put into quantum physics context.
They appear also  to be a link to statistical physics.
The following
theorem shows that the dynamic system $(\triangle_{n-1},M)$ has an unique
property iff $M$ is doubly stochastic. I will use a family of entropies, so
called $\alpha $-entropies, $\alpha >0$, defined by the formula:
\be
H_{\alpha }(p):=\frac{1}{1-\alpha } \ln ( \sum_i p_i^{\alpha } )
\ee
for $p\in \triangle_n$
The Shannon entropy appears for the limit $\alpha\rightarrow 1$,
$H_{\infty}=-\ln \max \{ p_i\}$,
for quantum $\alpha $-entropies see e.g. \cite{Th80}.

\begin{tm}   \label{second}
(a) For the dynamic system $(\triangle _{n-1},M)$ governing by the
stochastic matrix $M$ and for $\alpha \ge 1$ or $\alpha=\infty$:
$M$ is doubly stochastic iff the entropy $H_{\alpha}(M^kp)$
does not decreases for each trajectory. \\
(b) For generic
doubly stochastic matrices (for a dense and open set)
the entropy increases for all trajectories with
only one exclusion, the unique stationary trajectory. Then also
$\lim_{k\rightarrow \infty} M^kp=p_e$, where $p\in \triangle_{n-1},
p_e^i=\frac{1}{n}$.
\end{tm}

\begin{proof}
(a) For $M$ being a doubly stochastic matrix I will show that
$||M||_{\alpha }=1$ for $\alpha>1$, where
$||v||_{\alpha}:=\sqrt[\alpha]{\sum_i |v_i|^{\alpha}}$, $v\in {\bbold R}^n$,
and $||A||_{\alpha }:=\sup_{||v||_{\alpha }=1}||Av||_{\alpha }$. Each doubly stochastic matrix has the form:
$M=a_iP_i$, where $a_i\ge 0,\ \sum_i a_i=1$ and $P_i$ are the permutations. Now, $||M||_\alpha\le a_i||P_i||_\alpha=1$, where one uses $||P_i||_\alpha=1$ for each $i$.

But the condition, i.e. $||Mv||_\alpha\le ||v||_\alpha $, coincides with $H_{\alpha}(p)\le H_{\alpha}(Mp)$.
Inversely, it is enough to use a characterization of doubly stochastic
matrices through the equation
\be     \label{c}
Mp_e=p_e
\ee
While $H_{\alpha}$ does not decrease $p_e$ the point of reaching
the greatest value is not moved by $M$. \\

(b) By taking limit of $||Mv||_{\alpha}\le ||v||_{\alpha}$ for $\alpha
\rightarrow \infty $, where $M$ is a doubly stochastic matrix,
one concludes that $||M||=1$. The generic $M$ is defined as
$||Mp||<1$ for $p\in\triangle_{n-1}, p\ne p_e$.
Then, one estimates that
$||Mp-p_e||=||M(p-p_e)||<||p-p_e|| $ for $p\ne p_e$
and the proof is completed.
\end{proof}
\begin{rk}
The result is also kept for $\alpha=1$ and $\alpha=\infty$.
The extension for $\alpha=1$ contains von Neumann inequality
and its generalization \cite{Da91}. The theorem may be directly extended to
quantum case and infinite dimensional spaces with help of \cite{Sc00,Th80}.
\end{rk}
%

Returning to the beginning evolution proposal one may note that theorem
\ref{second} is immediately generalized for a dynamics constructing with
any chain of doubly stochastic matrices. For generic one it is that
$||M_1|_0\circ M_2|_0\circ \ldots ||_{\alpha}
\le ||M_1|_0||_{\alpha}\cdot ||M_2|_0||_{\alpha}\cdot \ldots =0$, where $M|_0$
is restriction of stochastic matrix $M$ to the subspace
$\{v\in {\bbold R}^n|\sum_i v^i=0 \}$.

Such abstract framework without all information about background process
appears to be powerful to explain some unsolved points connected with the
generalized second law of thermodynamics in black holes spacetimes
\cite{Be99}, also \cite{Be73,BCH73,Be74,Ha75}. I formally adopt the
system-bath model from condensed matter physics \cite{LBW00} on ${\cal H}_s
\otimes {\cal H}_b $ Hilbert space, where ${\cal H}_s$ is a space of states
of black holes and their neighborhood, ${\cal H}_b$ - of the bath (surrounding).
The only assumption I need is thermodynamic equilibrium of the bath.
The assumption is implicit present in the construction of dynamics via
Kraus representations. After a reduction to pure states one switches on
evolution with the same beginning vector state of the bath, the identity. Additionally,
the set of black holes, even gluing one with another in time, and
radiation and particles around are treated in quantum field theory fashion
as a quantum being with one Hilbert space.
Then ${\cal H}_s\otimes {\cal H}_b$
may be treated as a constant in time arena for quantum evolution. No
knowledge about theory of quantum dynamics of black holes is needed. Now,
the main result extended to the stochastic quantum
evolution is read as a
support of the generalized second law of thermodynamics admitting black holes
as subsystems. The environment presence, the last element needed
in Patrovi model, may be connected with structure of infinity of space-time
bearing during evolution some constants like ADM mass, charge or angular
momentum, a macroscopic parameters of inner quantum world.

The direct trial of finding a differential equation for $M(t)$ from the
\linebreak Schr\"odinger equation or its generalization
in the presence of the bath is
unsuccessful. There are no differential equation of the form
\be
M^{(n)}(t)=f_{\hat{H} }(t,M(t),\ldots ,M^{(n-1)}(t) )
\ee
where $\hat{H}$ is a
hamiltonian. For a comparison with
an approximation by the constant coefficients Lindblat equation see ref.
\cite{LBW00}. Really, it appears that the proposed model
leads to a non-Markovian evolution equation \cite{St01b}.

To summarize it needs to be stressed that theorem \ref{second} precisely
states that invertible quantum dynamics is after the pointing viewed as
a different side of entropy nondecreasing law.

The   approach   is  closely  related  to  the
consistent  history  formulation of Quantum Theory (QH) of Griffiths, Gell-Mann
and   Hartle   \cite{GMH93},   see  also  \cite{Is93,  Is96a,  Is96b}, its methods
and motivations.  The
above sequence  of  stochastic  matrices  is  equivalent  to a
consistent  histories  set  with  a  given  choice  of
orthogonal, complete, projective,   $1$-dimensional  operators  for  all  pointed  time  moments.
From the presented results follow that the unitary  evolution in QH may be
replaced  by  the  most general transformation for a
density  operator  of  a quantum  subsystem in thermal bath: $\hat{\rho}
\rightarrow \hat{\rho'}=
\sum_{\alpha}   A_{\alpha}\hat{\rho} A_{\alpha}^*$,   where
$
\sum_{\alpha}
A_{\alpha}A_{\alpha}^*=\sum_{\alpha} A_{\alpha}^*A_{\alpha}=1.
$
The theorem is directly rewritten
for any separable Hilbert space with the $\alpha $-entropies defined by
\be
H_\alpha(\hat{\rho})=\frac{1}{1-\alpha}Tr\hat{\rho}^{\alpha}
\ee
after identification of the eignvalues of $\hat{\rho}$ and $\hat{\rho'}$ with
$p, p'\in\triangle_{n-1}$, $n=1,2,\ldots,\infty $.
The reduction moments $t_i$ are admitted to possess wider gates of reduction, i.e. the complete, orthogonal set of projective operators $\{\hat{P_l(t_i)} \}$ i.e. $\sum_kP_k(t_i)^2=1$, $P_k(t_i)^{*}=P_k(t_i)$. The measurement is a discontinuous moment
evolution, but still realized by the doubly sum Kraus representation. It  may be even replaced by any doubly sum transformation.

\vspace{.5cm}
\noindent
{\bf Acknowledgments:} I would like to thank a referee for careful
reading the manuscript, detailed remarks and pointing related new
results. Also, I thank very much the Library of Physics Faculty of
University and the Main Library of Technology University in Warsaw
for help in completing the bibliography.

\bibliographystyle{prsty}

\bibliography{spontv2}

\begin{thebibliography}{10}

\bibitem{tH99}
G. 't~Hooft, Class. Quant. Grav. {\bf 16},  3263  (1999), gr-qc/9903084v2.

\bibitem{Pa89}
M.~H. Patrovi, Phys. Lett. A {\bf 137},  445  (1989).

\bibitem{BH01}
D.~C. Brody and P. Hughston, e-preprint  (2001), quant-ph/0011125.

\bibitem{LBW00}
D.~A. Lindar, Z. Bihary, and K.~B. Whaley, e-preprint  (2000),
  cond-math/0011204.

\bibitem{Sc00}
R. Schrader, e-preprint  (2000), math-phys/0007020.

\bibitem{AS97}
A. Ashtekar and T.~A. Schilling, e-preprint  (1997), gr-qc/9706069.

\bibitem{Lo97}
J.~D. Louck, Found. Phys. {\bf 27},  1085  (1997).

\bibitem{Th80}
W. Thirring, {\em {Lehrbuch der Mathematischen Physik}} (Springer-Verlag, Wien,
  1980), Vol.~4.

\bibitem{Da91}
J. Daboul, Phys. Lett. A {\bf 159},  213  (1991).

\bibitem{Be99}
J.~D. Bekenstein, e-preprint  (1999), gr-qc/9906058.

\bibitem{Be73}
J.~D. Bekenstein, Phys. Rev. {\bf D7},  2333  (1973).

\bibitem{BCH73}
J.~M. Bardeen, B. Carter, and S.~W. Hawking, Commun. Math. Phys. {\bf 31},  161
   (1973).

\bibitem{Be74}
J.~D. Bekenstein, Phys. Rev. {\bf D9},  3292  (1974).

\bibitem{Ha75}
S.~W. Hawking, Commun. Math. Phys. {\bf 43},  199  (1975).

\bibitem{St01b}
J. Stry\l{}a,   (2002), submitted to J. Math. Phys., e-preprint,
  quant-ph/0204160.

\bibitem{GMH93}
M. Gell-Mann and J.~B. Hartle,  in {\em {Proceedings of the NATO Workshop on
  the Physical Origin of Time Asymmetry, Mazagon, Spain, September 30 - October
  4, 1991}}, edited by J. Halliwell, J. Perez-Mercader, and W. Zurek (Cambridge
  University Press, Cambridge, 1993), e-preprint gr-qc/9304023.

\bibitem{Is93}
C.~J. Isham, J. Math. Phys. {\bf 35},  2157  (1994).

\bibitem{Is96a}
C.~J. Isham, Internat. J. Theoret. Phys. {\bf 36},  785  (1997).

\bibitem{Is96b}
C.~J. Isham, e-preprint  (1996), quant-ph/9612035.

\end{thebibliography}

\end{document}